# Pure skin effect obeying power partition in directed graphs

Wenwen Liu[1], Bumki Min[2]*, Shuang Zhang[1]*

[1]New Cornerstone Science Laboratory, Department of Physics, The University of Hong Kong, Hong Kong

[2]Korea Advanced Institute of Science and Technology

*Corresponding authors. Email: shuzhang@hku.hk; bmin@kaist.ac.kr

**Abstract**

**Non-Hermitian physics has received great attention recently. In particular, band structures in non-Hermitian systems can be engineered to exhibit various topological effects. Among them, one of the most intriguing phenomena is the non-Hermitian skin effect (NHSE). Here, we investigate NHSE in systems featuring directed chains or directed graphs, where the arrows denote the directions of the non-reciprocal hopping between neighbouring nodes. We show that the systems exhibit pure skin modes with non-oscillatory wavefunctions, in contrast to previously studied NHSE. Interestingly, the sum of the decay constants along different directions for each skin mode obeys a power partition rule, i.e. their sum is a fixed value and the value of each constant only depends on the ratio between the non-reciprocal hopping parameters and is independent of detailed graph configurations. Such Pure Skin Effect (PSE) can be explained by using a generalized method for solving the Generalized Brillouin-zone with multiple bulk states.**

In the quest to understand the fundamental principles that govern the behavior of physical systems, physicists have long been guided by the Hermitian nature of operators in quantum mechanics. However, recent strides in research have led to the exploration of a fascinating and unconventional domain known as non-Hermitian physics[1-3]. Departing from the familiar constraints of Hermitian operators, non-Hermitian physics opens doors to a rich tapestry of phenomena and concepts, including the emergence of complex eigenvalues[4], non-reciprocal wave propagation[5, 6], and exceptional points[7-9]. Such unconventional behavior has profound implications for our understanding of quantum mechanics[3, 10], acoustic systems[11, 12], and even in the realm of optics and photonics[13, 14].

The NHSE, a captivating manifestation of non-Hermitian physics, describes phenomenon that eigenstates are localized exponentially at the edge of a chain[15-21]. This effect, originated from nonreciprocal hopping and therefore exhibiting directional asymmetry in wave transmission, has the potential to redefine the landscape of unidirectional amplifier[22], energy harvesting[23], and signal processing technologies[24, 25]. Despite its profound implications, the wavefunctions are not perfectly decaying in an exponential way. Instead, the states are forming standing wave patterns with strong oscillatory features.

Here, we demonstrate a new family of non-Hermitian systems consisting of non-reciprocal coupling among a cluster of nodes that can form skin effect with non-oscillatory wavefunction what we called Pure Skin Effect (PSE). With uniform hopping parameters, the system can be viewed as directed chains and graphs. We further discover power partition of the decaying constants of the PSE modes along different directions. We also show that this PSE can be applied to higher dimensional non-reciprocal systems.

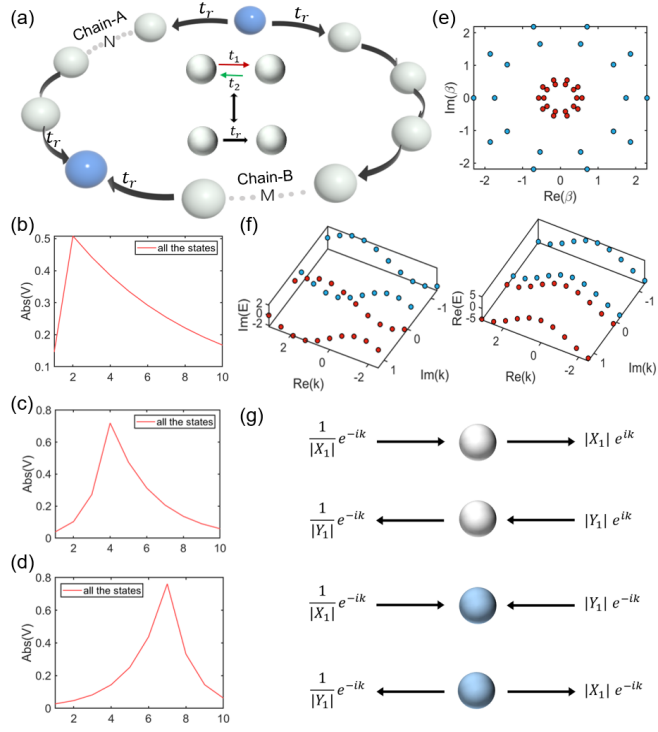

Fig. 1 (a) Schematic view of the general hopping model with two bulk chains of which the sites are represented by gray spheres. The blue spheres are the connecting points of the two chains. (b) Momentum distribution of the system with red (blue) dots represent the momentum of Chain-A (Chain-B). (c-e) Amplitude of wavefunction distribution along each site. Glowing and decaying coefficient is varied with the change of N and M. (f) Dispersion relation of both the real part and imaginary part of the energy (E) with respect to momentum (k). (g) Illustration of four different hopping situations that can happen to single sites in the general set up.

Here, we start with a directed chain loop consisting of two joined sections named as Chain-A and Chain-B with opposite hopping directions, and the corresponding number of hopping arrows in the clockwise and counter clockwise directions are N and M, respectively, as shown in Fig. 1(a). The two boundary sites are colored blue while other bulk sites are colored grey. The model can be mathematically described similar to HN model:

$$H = \sum_{i=1}^{N}(t_r c_i^\dagger c_{i+1} + c_{i+1}^\dagger c_i) + \sum_{i=N+1}^{N+M-1}(c_i^\dagger c_{i+1} + t_r c_{i+1}^\dagger c_i) + c_{N+M}^\dagger c_1 + t_r c_1^\dagger c_{N+M} \quad (1)$$

where $t_r = \frac{t_1}{t_2}$ represent the non-reciprocal hopping. As the hopping strength between two neighboring nodes are fixed to $t_1$ and $t_2$, only the direction of the hopping configurations matters and we could analogy the chain to a directed chain. The onsite energy is neglected in the sense that it is uniform among all the nodes. By giving exact number to the model, it is observed that the linked directed chain can always form PSE regardless the number of N and M. In Fig. 1(b-d), non-oscillatory wavefunction is observed when $[N, M] = [1, 9]$, $[N, M] = [3, 7]$ and $[N, M] = [6, 4]$, respectively, while it shows purely exponential profile between two boundary sites. Interestingly, a universal power partition rule is observed that the logarithm sum of decaying rate is conserved during the changing of $N$ and $M$ while $N + M$ is a constant.

To give the reason for the power partition phenomenon, we study the Generalized Brillouin-zone with multiple bulk states. The most general set up is given in the supplemental material and here we solve for the case with two chains thus having two bulk equations:

$$t_r \varphi_{n-1} - E\varphi_n + \varphi_{n+1} = 0$$
$$\psi_{n-1} - E\psi_n + t_r \psi_{n+1} = 0 \quad (2)$$

The two roots for the two sections are denoted as $\alpha_1$, $\alpha_2$ and $\beta_1$, $\beta_2$, respectively. The wave functions are the linear combination of the roots as $\varphi_n = C_1 \alpha_1^n + C_2 \alpha_2^n$ ($\psi_n = C_3 \beta_1^n + C_4 \beta_2^n$). Considering the boundary continuity, we have $M[C_1, C_2, C_3, C_4]^T = 0$, where:

$$M = \begin{bmatrix} \alpha_1^{N+1} & \alpha_2^{N+1} & -\beta_1 & -\beta_2 \\ \alpha_1 & \alpha_1 & -\beta_1^{M+1} & -\beta_2^{M+1} \\ \frac{1}{t_r}\alpha_1^2 & \frac{1}{t_r}\alpha_2^2 & -\beta_1^{M+2} & -\beta_2^{M+2} \\ \frac{1}{t_r}\alpha_1^{N+2} & \frac{1}{t_r}\alpha_2^{N+2} & -\beta_1^2 & -\beta_2^2 \end{bmatrix} \quad (3)$$

The eigenmodes and the corresponding momenta solution of the two bulk states $\alpha_1$, $\alpha_2$ and $\beta_1$, $\beta_2$. can be solved by setting $det[M] = 0$. The results are shown in Fig. 1(e), where the red and blue symbols represent the momentum of chain-A and Chain-B. As the chains have opposite direction, we get

$$M = \begin{bmatrix} \alpha_1^{N+1} & \frac{t_r^{N+1}}{\alpha_1} & -\frac{1}{\alpha_1} & -\frac{\alpha_1}{t_r} \\ \alpha_1 & \frac{t_r}{\alpha_1} & -\frac{1}{\alpha_1}^{M+1} & -\frac{\alpha_1^{M+1}}{t_r} \\ \frac{1}{t_r}\alpha_1^2 & t_r \frac{1}{\alpha_1}^2 & -\frac{1}{\alpha_1}^{M+2} & -\frac{\alpha_1^{M+2}}{t_r} \\ \frac{1}{t_r}\alpha_1^{N+2} & \frac{1}{t_r}(\frac{t_r}{\alpha_1})^{N+2} & -\frac{1}{\alpha_1}^2 & -\frac{\alpha_1^2}{t_r} \end{bmatrix} \quad (4)$$

From which analytical solution is obtained with $\alpha_1 = {}^{M+N}\sqrt{t_r{}^N} * e^{i\frac{2\pi n}{M+N}}$ using condition $DET(M) = 0$. Substitute again the solution to Eq. (4) and divide the third column $M(3)$ of the matrix by the first column $M(1)$, linear relation is observed as $M(3) = TM(1)$, in which $T$ is isomorphic to 4-dimension unit vector. This linear relative pair is always existing in the directed chain system. We prove in the supplemental material that when such kind of linear relative pair is here, we must have two of the combination coefficient C equals to zero. Recall that $\varphi_n = C_1\alpha_1{}^n + C_2\alpha_2{}^n$ and $\psi_n = C_3\beta_1{}^n + C_4\beta_2{}^n$. When $C_1(C_2)$ and $C_4$ $(C_3)$ equals to zero, $\varphi_n$ and $\psi_n$ becomes pure exponential function. Because $\alpha_1 = {}^{N+M}\sqrt{t_r{}^N} * e^{i\frac{2\pi m}{N+M}}$ and $\beta_2 = {}^{N+M}\sqrt{t_r{}^{-M}} * e^{i\frac{2\pi m}{N+M}}$, the decaying rate is $D = |\frac{\varphi_n}{\varphi_{n+1}}| = {}^{N+M}\sqrt{t_r{}^{-N}}$ and $G = |\frac{\psi_{n+1}}{\psi_n}| = {}^{N+M}\sqrt{t_r{}^{-M}}$. Therefore, we have $|\log_{t_r} D| + |\log_{t_r} G| = 1$. When changing N and M, the sum is not change while D and G change separately, which is actually the power partition rule. We also find another way to explain why the C equals to zero and show it in the supplementary material.

Now the mathematic derivation is clear, but we still wondering the physical meaning of this kind of wave propagation. Recall that the open boundary condition forms walls that cause the reflection. For our case, oppositely, the directed wave propagates smoothly through the boundary without reflection. Then it must have somewhere to go when travelling. Looking back to the bulk restriction Eq. (2), we can always define $\alpha_1 = r_1 e^{ik}$ and $\alpha_2 = r_2 e^{-ik}$, thus $\beta_1 = 1/r_1 e^{-ik}, \beta_2 = 1/r_2 e^{ik}$. According to the boundary restriction (Eq. (3)), $C_1\alpha_1 + C_2\alpha_2 = C_3\beta_1 + C_4\beta_2$, we have

$$C_1(r_1 e^{ik})^{N+1} + C_2(r_2 e^{-ik})^{N+1} = C_3/r_1 e^{-ik} + C_4/r_2 e^{ik}. \qquad (5)$$

Therefore, the momenta could smoothly transverse from $\alpha_1$ ($\alpha_2$) to $\beta_2(\beta_1)$ that makes k continue, namely, travelling in one direction. For one E, we only have unidirectional travelling wavefunction with single complex momentum that with the amplitude growing from one site to another and decaying when coming in a circle back (Fig. 1 (f)).

A more general directed chain is connecting a series of these two different chains [supplemental figure]. If the number of sites in each Chain-A is $[n_1, n_2 ... n_p]$ and the number of Chain-B is $[m_1, m_2 ... m_q]$, $p, q \in Z$, then the decaying rate would be

$\sqrt[n_1+\cdots+n_p+m_1+\cdots+m_q]{t_r^{-(n_1+\cdots+n_p)}}(\sqrt[n_1+\cdots+n_p+m_1+\cdots+m_q]{t_r^{-(m_1+\cdots+m_q)}})$. The amplitude of the wavefunction will decay with the same rate in all Chain-A and Chain-B area. To prove the casual combining case, we do reverse derivation as shown in Fig. 1 (g). For all these cases, we only have four different hopping structures. From the bulk equation, we have $E = t_1 \frac{\psi_{n-1}}{\psi_n} + t_2 \frac{\psi_{n+1}}{\psi_n}$, which means for one site in the middle, it can get power $\frac{\psi_{n-1}}{\psi_n}$ ($\frac{\psi_{n+1}}{\psi_n}$) from the left(right) with different hopping intensity ($t_1$ or $t_2$) in these cases. Therefore, we get $E = t_r^{\frac{m_1+\cdots+m_q}{n_1+\cdots+n_p+m_1+\cdots+m_q}} e^{-ik} + t_r^{\frac{n_1+\cdots+n_p}{n_1+\cdots+n_p+m_1+\cdots+m_q}} e^{ik}$ for all the cases. This means that the power partition can not only be power changed by different N and M but also be realized in any combination of the two chains. Another forward prof is shown in supplemental material.

By doing further calculation, we notice that the energy and momentum are all complex in our directed chain system. This is quite different from the conventional case in OBC. Figure 2 (a-c) show how the energy spectrum change with $\Delta = N - M$. When $\Delta \to 0$, the imaginary part of the energy decreases to zero. The energy become completely real which is similar to the open boundary case. It is easy to prove by using $E = (\alpha_1 + \alpha_2)/t_2$ and boundary restriction that $C_1 r_1^N = C_4/r_2 (C_2 r_1^N = C_3/r_2)$ and $C_1 r_1 = C_4/r_2^M (C_2 r_1 = C_3/r_2^M)$. To have E equals to zero, we should make $|\alpha_1| = |\alpha_2|$ that $r_1 = r_2$, which only happens when $N = M$. The corresponding eigenstate are also plotted in the inset of Fig 2(a-c). The oscillation happens abruptly when $\Delta = 0$. The reason is that the energy degeneracy happens when all the imaginary part of the energy becomes zero (Fig1(f)). As a result, the forward and backward propagating wave with the same energy meet each other and make the entanglement.

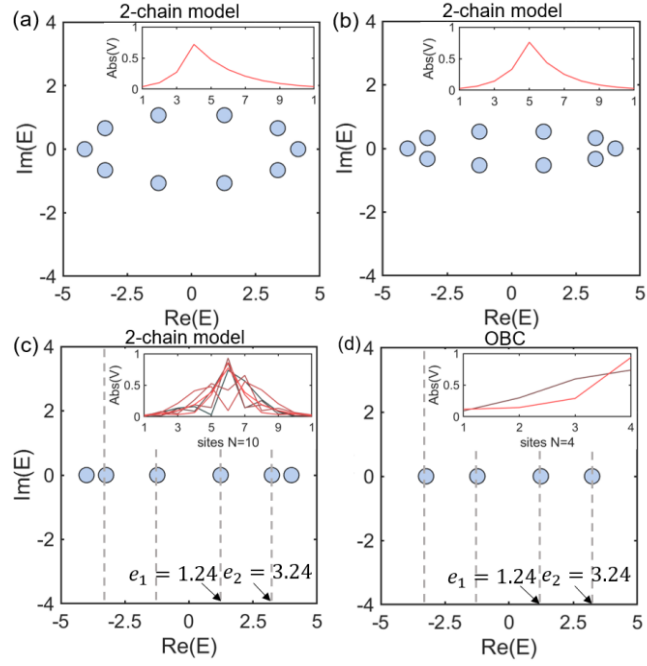

Fig. 2 (a-c) Energy spectrum of directed chain under different N and M but N+M=10. (d) Energy spectrum of conventional open boundary system with the bulk hopping is the same as either of the two chain and the site number is $\Gamma_2 = 4$.

As the $N = M$ case is quite similar to the conventional NHSE in OBC, we also try to figure out how can our case make same energy value comparing with the conventional one. We find that if the total site number of our system $\Gamma_1$ and the number of OBC $\Gamma_2$ have the relation $2\Gamma_2 + 2 = \Gamma_1$, the energy spectrum could have some value overlapping. The equation can be explained by driving analytical solution of the momentum of OBC by using same method of our general calculation method illustrated before. The resulting momentum $Z\pm =$ $^{2N_2+2}\sqrt{t_r}e^{\pm i\frac{2\pi m}{2N_2+2}}$ (see supplemental material) will be the same with $\alpha_1 = {}^{N+M}\sqrt{t_r{}^N} * e^{i\frac{2\pi m}{N+M}}$ if $N + M = \Gamma_1 = 2\Gamma_2 + 2$. However, it is worth mentioning that this oscillation is still different from that of the OBC as it is not formed by reflection. Therefore, the combining coefficient of two propagating wave is arbitrary instead of the same.

The PSE is also observed in well-designed directed charts which have more than two arrows linking to one site what we call channels. Sticking the set up to $t_r$ only, and reconstruct the chain structure, we get the chart shown in Fig 3(a-b). The corresponding wave function is plotted in the insert that also represent perfect exponential decaying from one site to the other without oscillation. The structure is more like an information transmission chart that allows information to accumulate or decrease progressively in a smooth way. We summarize the Hamiltonian of this kind of set up as follows:

$$H = \begin{bmatrix} 0 & \underbrace{\dots t_1 \dots}_{m} & \underbrace{\dots 0 \dots}_{n} & \underbrace{\dots t_1 \dots}_{m} & \underbrace{\dots 0 \dots}_{n} & \underbrace{\dots t_1 \dots}_{m} & \dots & \underbrace{\dots t_1 \dots}_{m} \\ \underbrace{\dots t_2 \dots}_{m} & 0 & \underbrace{\dots t_1 \dots}_{m} & \underbrace{\dots 0 \dots}_{n} & \underbrace{\dots t_1 \dots}_{m} & \underbrace{\dots 0 \dots}_{n} & \dots & \underbrace{\dots 0 \dots}_{n} \\ \underbrace{\dots 0 \dots}_{n} & \underbrace{\dots t_2 \dots}_{m} & 0 & \underbrace{\dots t_1 \dots}_{m} & \underbrace{\dots 0 \dots}_{n} & \underbrace{\dots t_1 \dots}_{m} & \dots & \underbrace{\dots t_1 \dots}_{m} \\ \underbrace{\dots t_2 \dots}_{m} & \underbrace{\dots 0 \dots}_{n} & \underbrace{\dots t_2 \dots}_{m} & 0 & \underbrace{\dots t_1 \dots}_{m} & \underbrace{\dots 0 \dots}_{n} & \dots & \underbrace{\dots 0 \dots}_{n} \\ \underbrace{\dots 0 \dots}_{n} & \underbrace{\dots t_2 \dots}_{m} & \underbrace{\dots 0 \dots}_{n} & \underbrace{\dots t_2 \dots}_{m} & 0 & \underbrace{\dots t_1 \dots}_{m} & \dots & \underbrace{\dots t_1 \dots}_{m} \\ \underbrace{\dots t_2 \dots}_{m} & \underbrace{\dots 0 \dots}_{n} & \underbrace{\dots t_2 \dots}_{m} & \underbrace{\dots 0 \dots}_{n} & \underbrace{\dots t_2 \dots}_{m} & 0 & \dots & \underbrace{\dots 0 \dots}_{n} \\ \dots & \dots & \dots & \dots & \dots & \dots & \ddots & \underbrace{\dots t_1 \dots}_{m} \\ \underbrace{\dots t_2 \dots}_{m} & \underbrace{\dots 0 \dots}_{n} & \underbrace{\dots t_2 \dots}_{m} & \underbrace{\dots 0 \dots}_{n} & \underbrace{\dots t_2 \dots}_{m} & \underbrace{\dots 0 \dots}_{n} & \underbrace{\dots t_2 \dots}_{m} & 0 \end{bmatrix} \quad (6)$$

in which $n, m \in Z$. No matter what the number of $n$ and $m$ is, the Hamiltonian can always form a recursive sequence if we decompose the matrix column by column to equations and subtract them one by one. This mathematically allows the exponential decaying with the increase of the sequence (supplemental). It is also easy to prove that the recursive sequence is naturally formed in 1D directed chain system. This theory generally allows us to generate such kind of information transmission in very rich configurations.

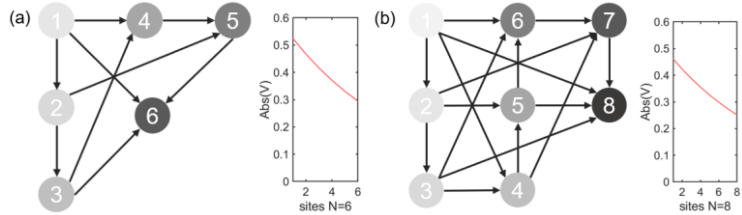

Fig. 3 Directed chart that support PSE. (a) Directed chart structure (left) and corresponding states distribution (right) with site number $N = 6$ and channel number $C = 3$. (b) Directed chart structure (left) and corresponding states distribution (right) with $N = 8$ and $C = 4$.

Now we have discussed the cases within the system that only have one hopping coefficient with various directions which can all be classified to 1D cases. The theory can also be expanded to 2D system as shown in Fig 4(a). Instead of forming a circle or building chart by arranging arrow's direction with same $t_r$, we built 2D lattice by cascading the 1D chains or charts with different $t_r$ or structures in an orthogonal way. Along each orthogonal direction, the charts are the same and cascading itself can form the 1D directed chains or charts. Fig 4(a) shows the general set up of the combination of two directed chains, which is analog to a torus established by folding the 2D lattice bottom to top and left to right, like 2D Brillouin Zone. All the chains in the horizontal or vertical direction can also be replaced by same directed chart as shown in the insert. As they are orthogonal to each other, the wave propagating along horizontal and vertical direction do not interfere with each other, which allows us to adjust them separately. We use $t_{r1}$ and $t_{r2}$ to distinct the different hopping coefficient.

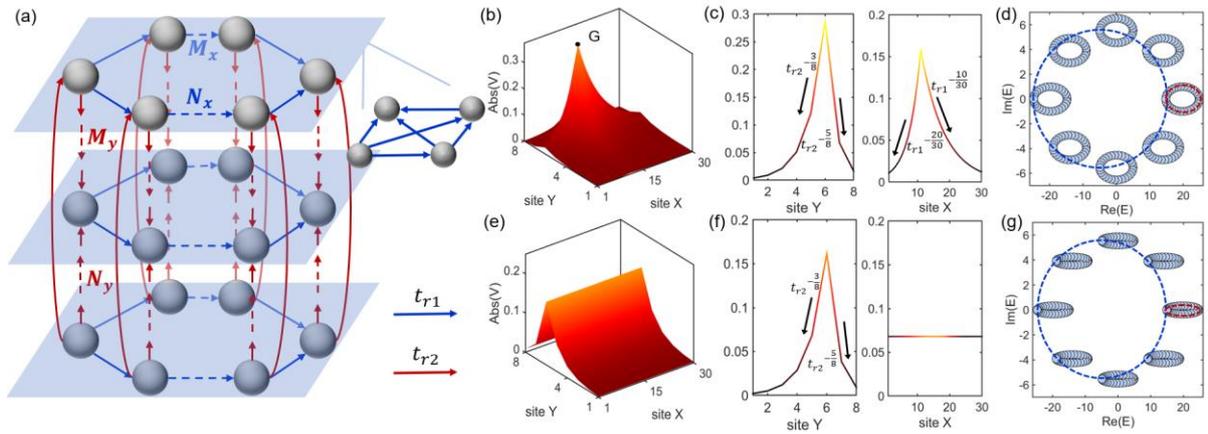

Fig. 4 (a) Schematic view of the expanded directed chains in 2D system. $N_x$, $M_x$ and $N_y$, $M_y$ represent the number of sites in Chain-A and Chain-B with hopping $t_{r1}$(blue), Chain-C and Chain-D with hopping $t_{r2}$ (red). The chains in the blue plane can be changed to any directed charts as shown in the insert. (b-c) Amplitude distribution when $N_x=10, M_x = 20, N_y = 5, M_y = 3, t_{r1} = 1.5, t_{r2} = 10$ along each site on 2D view (b) and cross section view (c). (d) Energy spectrum corresponding to (b), the (red) blue circle and red circle represent the energy corresponding to same $k_y$ ($k_x$). (e-f) Amplitude distribution when $N_x=30, M_x = 0, N_y = 5, M_y = 3, t_{r1} = 1.5, t_{r2} = 10$ along each site on 2D view (b) and cross section view (e). (g) Energy spectrum corresponding to (e).

Fig. 4(b) shows the result of setting $N_x=10, M_x = 20, N_y = 5, M_y = 3$ for the two-chain case. A bright corner state is formed in the Q point. There is also a minimum value state localized at the site $x = 0$ and $y = 0$ point. We give the cross section view with site $x = 10$ and site $y = 5$(Fig. 4(c)). It is clearly seen that the decaying rate along y direction is $t_{r2}^{-\frac{3}{8}}(t_{r2}^{-\frac{5}{8}})$ and along x direction is $t_{r1}^{-\frac{20}{30}}(t_{r1}^{-\frac{10}{30}})$ , which intuitively shows that the two directions are independently controlled. We can also form edge state by setting one direction hopping to be periodic. The result is shown in Fig 4 (e-f) which shows perfect edge state along x direction.

The phenomena can be understood by looking into the energy spectrum (Fig. 4 (d, g)). The energy form $N_x + M_x$ big loops and $N_y + M_y$ small loops. For the point on the small loop circled by red (blue) line, they have same $k_y$ ($k_x$) for each $k_x$ ($k_y$), which will correspond to single unidirectional propagating wave. The perfect propagation will retain as long as the big loop and small loop do not collapse into a line or overlap with each other. In addition, power partition is realized in both dimensions as marked in Fig 4 (c, f). This theory can be extent to even higher dimension as shown in supplemental material.

Such symmetry breaking allowed no fluctuations information transfer is a general set up that can also be applied to Hermitian system with non-reciprocal hopping or non-Hermitian system with imaginary hopping terms. For such case, instead, there will not be amplitude increase or decrease but phase shift which influence the propagating characteristics. Non-oscillated wave is also observed in all this cases and more details can be found in supplemental material.

To conclude, we proposed the linked directed chain structure that can always realize non-oscillated PSE. All the eigenstate represents the special decaying form with certain exponential constant. Any rate of the power partition is realized by differing the number of sites in two opposite directed chains. The theory can be expanded to directed chart, 2-dimension which support edge states and corner states or system with unbalanced imaginary hopping terms. Even higher dimension is also allowed. Potential applications can be found in more efficient amplifiers and information transport.